\documentclass[twocolumn,showpacs,preprintnumbers,amsmath,amssymb,aps,prb,superscriptaddress]{revtex4-1}

\pdfoutput=1
\usepackage[pdftex]{graphicx}
\usepackage[english]{babel}

\usepackage{cleveref}

\usepackage{dcolumn}
\usepackage{bm}

\usepackage{color}
\usepackage{amstext}
\usepackage{braket}

\begin{document}


\title{In-plane magnetic field \textit{vs}. temperature phase diagram of a quasi-2D frustrated multiband superconductor}

\author{A. M. Marques}
\affiliation{Department of Physics $\&$ I3N, University of Aveiro, 3810-193 Aveiro, Portugal}
\author{ R. G. Dias}
\email{rdias@ua.pt}
\affiliation{Department of Physics $\&$ I3N, University of Aveiro, 3810-193 Aveiro, Portugal}
\author{ M. A. N. Ara\'ujo}
\affiliation{CFIF, Instituto Superior T\'ecnico, University of Lisbon, Avenida Rovisco Pais, P-1049-001 Lisboa, Portugal}
\affiliation{Department of Physics, University of \'Evora, P-7000-671 \'Evora, Portugal}
\author{ F. D. R. Santos}
\affiliation{Department of Physics $\&$ I3N, University of Aveiro, 3810-193 Aveiro, Portugal}
\date{\today}


\begin{abstract}

Motivated by the recent discovery of iron-based superconductors, with high critical temperatures and multiple bands crossing the Fermi level, we address the conditions for the presence of chiral superconducting phases configurations in the in-plane magnetic field \textit{vs}. temperature phase diagram of a quasi-2D frustrated three-band superconductor. 
Due to Zeeman splitting, the coupled superconducting gap equations present a complex set of solutions. For weak interband couplings, chiral configurations are only attained in a narrow strip of the in-plane magnetic field \textit{vs}. temperature phase diagram. This strip of chiral states becomes narrower and disappears at low temperatures, giving way to a first-order transition between non-chiral superconducting states. For stronger interband couplings, the chiral strip is much broader, if the interband couplings are approximately equal; otherwise, the chiral region is expected to be completely absent of the phase diagram.
\end{abstract}

\maketitle

\section{Introduction}
\label{sec:intro}
Recently, the possibility of chiral states with broken time-reversal symmetry in  a multiband superconductor has been discussed \cite{NgT.K.2009,Stanev2010,Tanaka2010,Tanaka2010a,Dias2011,Garaud2011,Hu2012,Wilson2013,Orlova2013,Gillis2014}. In a multiband superconductor, interband pairings may be regarded as  internal Josephson tunnellings \cite{Leggett1966,Agterberg2002,Moll2014}, and  repulsive  interband interactions (proposed to be present, for instance, in sign-reversed  iron-based superconductors \cite{Kuroki2008,Mazin2008}) are the equivalent of $\pi$-junctions in a Josephson junction array, since a  $\pi$-junction is also characterized by a sign change (a $\pi$ phase shift) in the order parameters across the junction \cite{Tsuei2000b,Li2003}. A  $\pi$-junction can be realized in a number of ways, such as: by inserting a thin ferromagnetic layer \cite{Ryazanov2001,Kontos2002} or an insulating oxide layer with magnetic impurities \cite{Kulik1965,Bulaevskii1977} between superconductors, or even by growing grains of $d$-wave superconductors, typically hole or electron-doped high-$T_c$ cuprate superconductors, with a $45^{\circ}$ misorientation in the c-axis at the grain boundaries \cite{Kouznetsov1997,Schulz2000, Tsuei2000, Lombardi2002,Chesca2003}.
While for two-band superconductors free energy minimization requires that the superconducting phases are either aligned or anti-aligned, the situation changes if three or more bands and an odd number of repulsive interband interactions are present \cite{Dias2011}.
In this frustrated case, the stable phase configuration depends on the relative strengths of the couplings involved and three-band superconductors may lock their phases in values that depart from $0$ or $\pi$, producing chiral configurations. Variation of temperature may cause second-order transitions from or to chiral regions. By tuning the coupling parameters and controlling the temperature, one can manipulate the three-band superconductor in order to stabilize virtually in any given phase configuration.

On the experimental side, some iron-based compounds have been reported to have three or more bands participating in superconductivity \cite{Nakayama2009,Zhang2010,Storey2013}, making them the most likely candidates to exhibit intrinsic frustration effects. Due to their  reduced dimensionality, iron-based superconductors show reduced orbital effects when in-plane magnetic fields are applied. Under such magnetic fields, Zeeman splitting becomes the dominant pair breaking factor, and as a first approximation the orbital effects can be neglected.
In-plane magnetic fields influence the values of the superconducting gaps and provide another way to change the relative superconducting phases of the bands. In the in-plane magnetic field \textit{vs.} temperature phase diagram of one-band superconductors, a curve of first-order phase transitions with increasing magnetic field is known to be present at low temperatures, ending at a tricritical point around $T^\star \approx 0.56 T_c$. For $T>T^\star$ the transition becomes of second-order \cite{Maki1964,Shimahara1994}. The paramagnetic limit $H_p=\mu \Delta_0/\sqrt{2}$, also
designated as Pauli limit or Chandrasekhar-Clogston limit \cite{Chandrasekhar1962,Clogston1962},
is the zero-temperature critical magnetic field associated with a first-order transition (FOT). 
In quasi-2D two-band superconductors with weak interband interactions, besides the above mentioned FOT curve, an additional low temperature FOT curve appears within the
superconducting region of the phase diagram, characterized by a large reduction of the superconducting gap in one of the bands \cite{Dias2005}.
As the interband coupling grows, this transition within the superconducting region approaches the FOT to the normal phase and disappears for a strong enough interband coupling.
In the case of a n-band superconductor, a cascade of FOTs curves is expected at low temperatures and weak interband couplings. However, for superconductors with $n>2$ bands and an odd number of repulsive interband couplings, one has also to allow for the possibility of having transitions to or from regions of chiral configurations of the superconducting phases\cite{Dias2011}.
The in-plane magnetic field versus temperature phase diagram of these superconductors becomes therefore more complex with the emergence of regions of chirality.



In this paper, we address the presence of chiral superconducting phases in the in-plane magnetic field \textit{vs.} temperature phase diagram of quasi-2D frustrated three-band superconductors in the weak and strong interband couplings regimes. For weakly coupled bands, the chiral region of the diagram consists only of a narrow strip, which becomes wider as the interband couplings are increased, eventually occupying most of the superconducting region of the phase diagram in the case of strongly coupled bands. We also address the magnetic field evolution of the superconducting phases and gap functions for systems with weak and intermediate interband couplings, and study the interplay between the chiral solutions for the phases and the FOTs in the superconducting state. Finally, we will consider how an asymmetry in the magnitudes of the interband couplings affects the possibility of finding chiral phase configurations and FOTs in the superconducitng state.

The remaining part of this paper is organized in the following way. In section \ref{sec:af}, we show how to determine solutions corresponding to chiral and non-chiral superconducting phase configurations for a quasi-2D three-band superconductor with one repulsive interband interaction. In section \ref{sec:weak}, we discuss the in-plane magnetic field \textit{vs.} temperature phase diagram for a frustrated system of three superconducting bands both in the weak and in the strong interband couplings regime. In section \ref{sec:stable} we study the magnetic field evolution of the superconducting gaps, superconducting phases and free energy difference between normal and superconducting states for three different temperatures for the case of weakly coupled bands. Also in section \ref{sec:stable}, the same studies were carried out for different frustrated three-band superconductors with intermediate interband couplings, and in particular considered one them to have an asymmetry in the magnitudes of the interband couplings. We conclude in section \ref{sec:conclusion}. The formalism to determine the free energy difference between normal and superconducting states for multiband superconductors is given in Appendix \ref{sec:appendix}.

%
%
%
%

\section{Hamiltonian and free energy}\label{sec:af}

We start by considering the BCS Hamiltonian for a $n$-band quasi-2D superconductor, including a Zeeman splitting term corresponding to the dominant effect of an applied in-plane magnetic field,
\begin{eqnarray}
    \mathcal{H}-\mu N-\sigma hN 
    &=&\sum_{\textbf{k}\sigma i} \xi_{\textbf{k}\sigma i}c_{\textbf{k}\sigma i}^\dagger c_{\textbf{k}\sigma i}   \nonumber  \\
    &-& \sum_{\textbf{kk}^\prime ij}V_{\textbf{kk}^\prime}^{ij}
    c_{\textbf{k}\uparrow j}^\dagger c_{-\textbf{k}\downarrow j}^\dagger 
    c_{\textbf{k}^\prime \uparrow i}c_{-\textbf{k}^\prime \downarrow i}
    \label{eq:hamiltonian} 
\end{eqnarray}
where $\xi_{k\sigma i}=\varepsilon_{ki} - \mu - \sigma h$, $\mu$ is the chemical potential, $\sigma=\uparrow,\downarrow$ is the spin component along the in-plane magnetic field, $h=\mu_BH$ and $\mu_B$ and $H$ are the Bohr magneton and the in-plane applied magnetic field, respectively. 
Superconductivity  phases $\phi_i$ associated with the superconducting parameters $\Psi_i=\braket{c_{k \uparrow i}c_{-k \downarrow i}}= \delta_i e^ {i \phi_i}$, with $\delta_i$ real, only affect the interband contributions for the free energy. Writing the energy in the mean field approximation of (1) in terms of these $\Psi_i$ yields \cite{Dias2011}
\begin{eqnarray}
E &=&    \sum _{i}f_i(\vert \Psi_i \vert^2)
      -\sum_iV_{ii}\vert \Psi_i \vert^2-
      \sum_{i\neq j}V_{ij}\Psi_i^{\star}\Psi_j  \nonumber
\\
  &=&    \sum _{i}f_i(\delta_i^{2})
      -\sum_iV_{ii}\delta_i^{2}-
      \sum_{i > j}J_{ij}\cos(\phi_{j}-\phi_{i}) ,
\end{eqnarray} 
where $f_i(\delta_i^{2})$ is the kinetic energy contribution of the respective band (given in Appendix A) and the two last terms give the intraband and interband coupling terms, respectively. $J_{ij}=2V_{ij}\delta_i\delta_j$ is the effective Josephson interband coupling. Note that only the interband term is phase-dependent.
The minimization of this energy with respect to the phases $\phi_{i}$  and to the absolute values of the superconducting parameters \cite{Shimahara1994,Leggett1966,Dias2011} yields a set of coupled equations for the gap functions,
\begin{eqnarray}
\Delta_i &=& \sum_{j}\cos(\phi_{j}-\phi_{i})  V_{ij}\delta_j,
         \label{eq:Deltas}
         \\
\delta_j &=& \int_0^{\hbar\omega_D} d\xi K_j \Delta_j,
\label{eq:deltas}
\end{eqnarray}
where
\begin{eqnarray}
          K_j  &=& K_j(\xi,\Delta,T,h)  \label{eq:kernel}\\
          &=& \dfrac{N_j(\xi)}{2E}
          \left( \tanh \frac{E+h}{2k_B T}+\tanh  \frac{E-h}{2k_B T} \right) 
          \nonumber,
\end{eqnarray}
and  where  $ E=\sqrt{\xi^2+\Delta^2}$, $\xi$ is the non-interacting energy dispersion, $\omega_D$ is the usual frequency cutoff and $N_j(\xi)$ is the density of states of band $j$. We assume  equal and constant density of states for all bands,  $N_j(\xi)=N_j(0)=N$. Differences in the density of states could also be absorbed in the couplings definition. The effect of the superconducting phase differences is the renormalization of the gap functions and, consequentially, of the effective Josephson interband couplings, given that $J_{ij}$ depends on the gap functions through (4). 

We will now restrict our study to three bands ($n=3$) but the arguments are easily generalized to arbitrary $n$.
We impose for a matter of convenience that $\phi_{1}=0$ (with no loss of generality). The explicit expression for the phase minimization of (2) is simply
\begin{eqnarray}
&\frac{\partial E}{\partial \phi_i}& = 0 \nonumber 
\\
\Rightarrow &\sum_j& J_{ij}\sin(\phi_j - \phi_i) = 0
\label{eq:minimization1}
\\
\Rightarrow &\sum_j& \sin(\phi_j - \phi_i)V_{ij}\delta_j = 0.
\label{eq:minimization}
\end{eqnarray}
Equations (\ref{eq:Deltas}) and (\ref{eq:minimization}) can therefore be merged into a more compact form,
\begin{equation}
\Delta_i = \sum_{j}e^{i(\phi_{j}-\phi_{i})}  V_{ij}\delta_j,
\end{equation}
where the condition of having all gap functions real is guaranteed by (\ref{eq:minimization}).
By solving (\ref{eq:minimization1}), we get several solutions corresponding to extreme or saddle points of the interband energy contribution, the non-chiral solutions being  $(\phi_1,\phi_2,\phi_3)=(0,0,0)$, $(0,\pi,0)$, $(0,0,\pi)$, and $(0,\pi,\pi)$ and the chiral solutions being \cite{Dias2011}
\begin{equation}
	(\phi_1,\phi_2,\phi_3)=\pm \left[0,\cos^{-1} (\alpha^-), -\text{sgn} \left(\frac{a}{b}\right) \cos^{-1}(\alpha^+) \right],
\end{equation}
where
\begin{equation}
	\alpha^{\pm}=\frac{\pm a^2 \mp b^2-a^2 b^2}{2 a b \gamma^{\pm}},
\end{equation}
$\gamma^{+}=b$, $\gamma^{-}=a$, $a=J_{12}/J_{23}$ and $b=J_{31}/J_{23}$. These chiral solutions exist only if  $\vert \alpha^{\pm} \vert \leq 1$. 



\begin{figure*}[t]
\begin{center}
\includegraphics[width=.8 \textwidth]{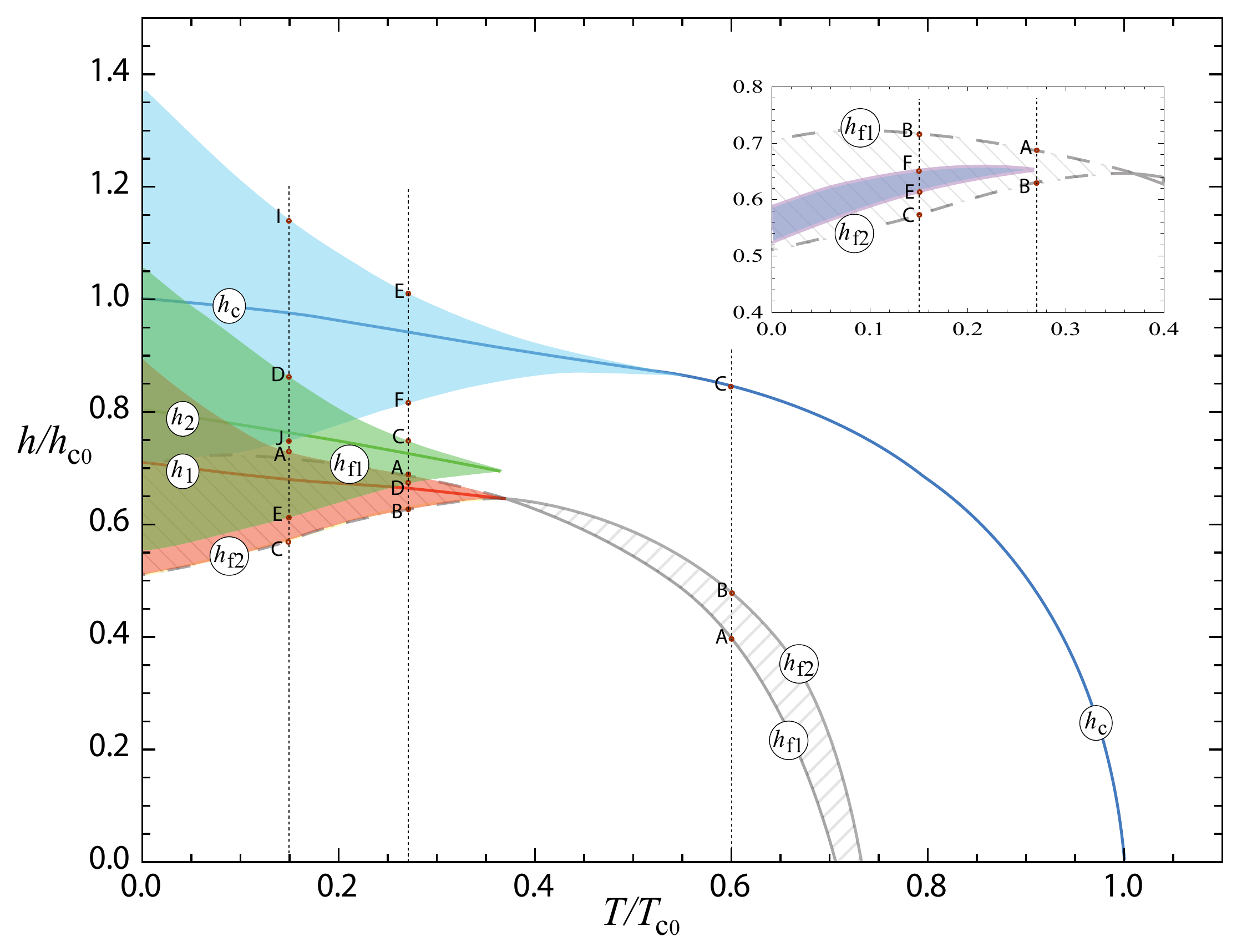}
\end{center}
\caption{In-plane magnetic field vs. temperature phase diagram for a system of three weakly coupled bands with one of the interband interactions being repulsive [see (\ref{eq:potential_matrix}) in the text]. Normalization values are $h_{c0}$, the thermodynamic critical field for zero temperature, and $T_{c0}$, the critical temperature for zero magnetic field. $h_1$, $h_2$, $h_c$ and $h_{f1,2}$ indicate, respectively, the first and second first-order transitions, the thermodynamic critical magnetic field and the magnetic fields that marks the boundaries of the chiral region (diagonal pattern), which interchange at the intersection at $T\approx 0.37T_{c0}$; for low temperatures this pattern has a negative slope to indicate that these chiral solutions fall into the metastable region of the $h_1$ transition. Shaded (red, green and blue) areas are regions of metastability associated with the first-order phase transitions and are limited by the corresponding supercooling field from below and by the superheating field from above. The dark (blue) area in the inset shows the existence of a second chiral region  within the first one. Vertical dashed lines with (red) dots and letters indicate the three temperatures shown in Fig.\ref{deltas}.}
\label{h_t_phase_diagram_1}
\end{figure*}


In Appendix A we derive the expression for the free energy difference between the superconducting and normal states,
\begin{eqnarray}
    \label{eq:freeenergy}
    F_s - F_n 
    &=&
    k_B T\sum_{\textbf{k}\sigma i} 
    \ln\frac{1-f(E_{\textbf{k}i}^\sigma)}{1-f(\vert \xi_{\textbf{k}i}^\sigma\vert)}  \nonumber
    \\
    &+&
    2\sum_{\vert \textbf{k} \vert >k_f,i}(\xi_{\textbf{k}i} - E_{\textbf{k}i}) + \sum_i \delta_i\Delta_i ,
    \label{free_energy}
\end{eqnarray}
where $E_{ki}^\sigma$ is the quasi-particle excitation energy of the superconducting state, $E_{ki}=E_{ki}^\sigma+\sigma h$ is the same energy in the absence of any applied magnetic field and $\vert \xi_{ki}^\sigma\vert=\vert \xi_{ki}\vert- \sigma h$ is the kinetic energy of a normal state electron in band $i$, in state $k$, with spin $\sigma$, measured from the Fermi energy. 
%

\section{Phase diagram for weak and strong interband couplings}\label{sec:weak}


We begin by considering a system of three weakly coupled bands where one of the interband couplings is repulsive (negative), to ensure the possibility of finding chiral states. The coupling parameters used, in terms of $V_{11}$, were the following
\begin{equation}
\begin{pmatrix}
V_{11} & V_{12} & V_{13}
\\
V_{12} & V_{22} & V_{23}
\\
V_{13} & V_{23} & V_{33}
\end{pmatrix}
=
V_{11}
\begin{pmatrix}
1 & -0.004 & 0.016
\\
-0.004 & 0.95 & 0.016
\\
0.016 & 0.016 & 0.88
\end{pmatrix}
.
\label{eq:potential_matrix}
\end{equation}

\begin{figure*}[t]
\begin{center}
\includegraphics[width=.8 \textwidth]{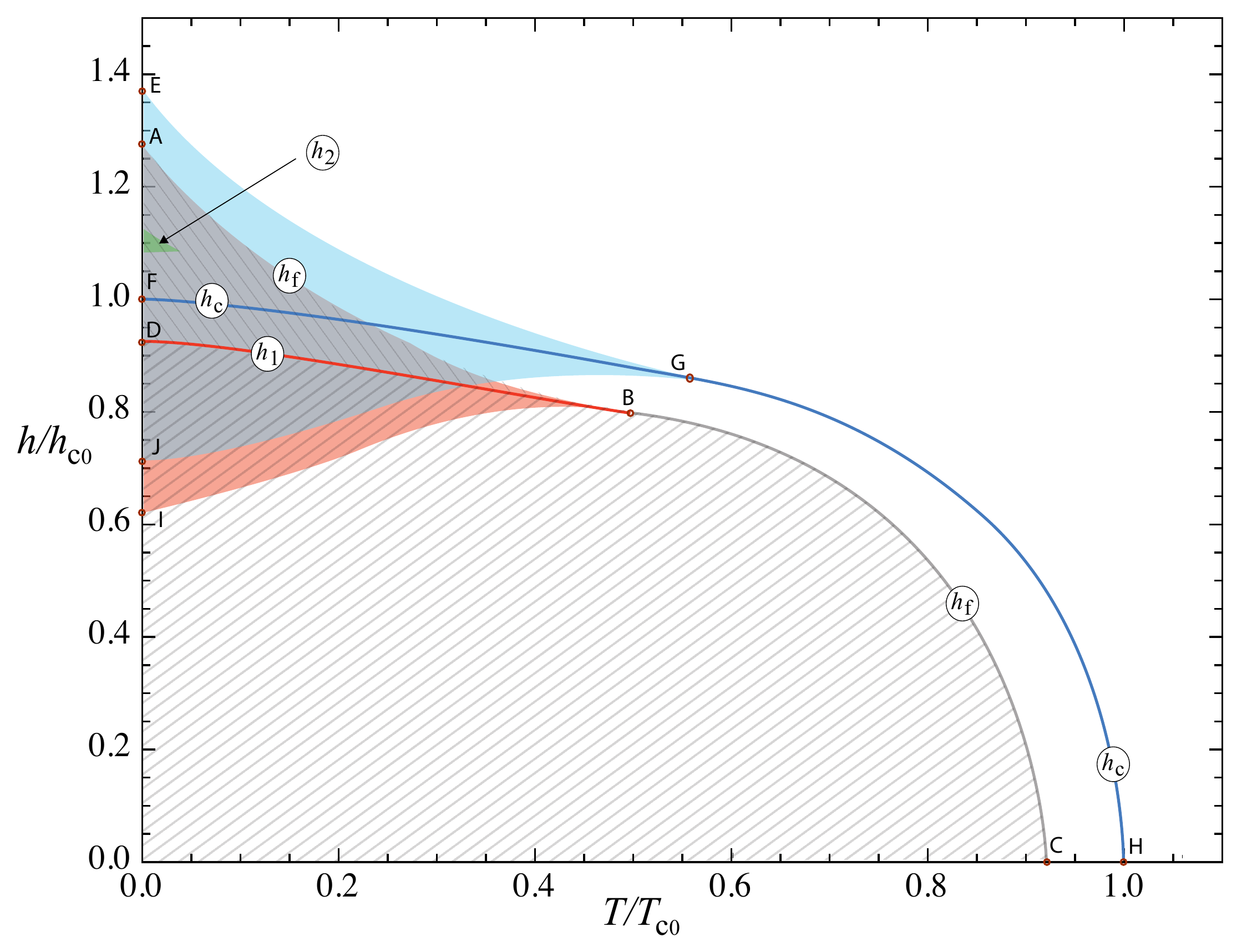}
\end{center}
\caption{In-plane magnetic field vs. temperature phase diagram for a system of three strongly coupled bands with one of the interband interactions being repulsive [see (\ref{eq:potential_matrix2}) in the text]. The notation is the same as in Fig.~\ref{h_t_phase_diagram_1}. The diagonal pattern in the ABD region has a negative slope to indicate that these chiral solutions are in the metastable region of $h_1$. The small (green) metastability region of the $h_2$ transition is present in the EGF region, that is, above the thermodynamic critical field $h_c$; at the very low temperatures where this metastable region is present, the behavior of the gap functions looks similar to the DE section in the left column of Fig. \ref{figures7}.}
\label{h_t_phase_diagram_2}
\end{figure*}

We used a self-consistent method to find the numerical solutions for the gap functions and their respective phases, from which the free energy difference was determined using (\ref{free_energy}). 
The results were condensed into the in-plane magnetic field versus temperature phase diagram of Fig.\ref{h_t_phase_diagram_1}. As expected there are now two FOTs within the superconducting phase at low temperatures ($T<0.37T_{c0}$), given by the (red) $h_1$ and (green) $h_2$ curves; in general, for $n$ weakly coupled bands one should find $n-1$ such transitions, located in the vicinity of the low temperature thermodynamical critical magnetic fields of the  bands with smaller gaps when they are considered uncoupled. Similarly, in the $\Delta$-$T$ plot of a multiband superconductor, one observes a sharp reduction of the smaller  gap functions  around their uncoupled critical temperatures (but remaining finite). The diagonal pattern, limited by the $h_{f1}$ and $h_{f2}$ curves, with positive (negative) slope indicates the region where stable (metastable) chiral phase configurations can be found. For $T \succsim 0.37T_{c0}$, the system always crosses the stable region of chirality with increasing field, and there is in fact a specific temperature interval where for zero field the system is already in a chiral phase configuration \cite{Dias2011}.  
With decreasing temperature, the point where the bottom FOT starts ($h_1$ curve), at $T\approx 0.37T_{c0}$, is also the point where the $h_{f1}$ and $h_{f2}$ curves are interchanged, making the $h_{f2}$ curve now coincide with the first supercooling field. In the next section, we will show how these chiral solutions at low temperatures ($T<0.37T_{c0}$) fall in the metastable region of the first transition, meaning that for low temperatures the stable phase solutions are always non-chiral. The smaller region of chirality that we see inside the larger one in the inset also corresponds to metastable solutions.

Next we consider a frustrated system of strongly coupled bands. The matrix of couplings, in terms of $V_{11}$, is given by
\begin{equation}
\begin{pmatrix}
1 & -4 & 4
\\
-4 & 0.95 & 4
\\
4 & 4 & 0.88
\end{pmatrix}
\label{eq:potential_matrix2}
\end{equation}
where now the interband couplings are approximately four times larger than the intraband couplings, which are kept the same as before (and throughout the paper).
This dominance of interband over the intraband pairing terms is expected to be relevant for example in iron pnictides \cite{Bang2008,Chubukov2008}.
The in-plane magnetic field versus temperature phase diagram relative to this case is shown in Fig. \ref{h_t_phase_diagram_2}. When compared with Fig. \ref{h_t_phase_diagram_1} it becomes clear that now the chiral region is much wider, occupying the DBCO region. Even though there are chiral solutions for the superconducting phases in the ABD region, these will not be observed since they are present in the metastable region of the first FOT: when $h_1$ occurs with increasing field, the system jumps to a stable non-chiral solution. In Fig.~\ref{figures7}(k) we show an example of this kind of transition from a chiral to a non-chiral superconducting phases configuration at the FOT (red arrows on the vertical dashed line) in the superconducting state. Therefore, in the entire superconducting region of the phase diagram, chiral solutions will only be absent in the small FGHCBD strip. 

Another difference relative to the weakly coupled system is that now we only have one FOT in the superconducting state. The $h_2$ curve is absent, but its metastability region is present above the $h_c$ transition to the normal state, in its larger metastable region. If we were to increase the interband potentials even more, we would expect the $h_2$ metastable region to disappear completely and the $h_1$ curve to follow the same general behaviour as $h_2$, that is, disappearing after crossing the critical $h_c$ curve and having its metastable region continuously rising and shrinking, and eventually disappearing. The evolution of the FOT curves in the superconducting phase in our three-band superconductors is consistent with the evolution of the single FOT in the superconducting phase of the two-band superconductors considered by Dias \citep{Dias2005}.

\section{Stable and metastable solutions of the coupled gap equations}\label{sec:stable}
In this section, we show the evolution of the coupled superconducting gaps, superconducting  phases and free energy difference between superconducting and normal states with applied magnetic field for specific values of temperature,
indicated by the vertical lines in the phase diagram  of  Fig.~\ref{h_t_phase_diagram_1}.

We also study the evolution of these parameters with applied magnetic field for systems with intermediate interband couplings, at low temperature. In the last case considered, we impose that one of the interband couplings is substantially different from the other two in magnitude, and we discuss the effects of this asymmetry.
\subsection{Weak interband couplings}
\begin{figure*}[t]
\begin{center}
\includegraphics[width=18.2cm,height=17cm]{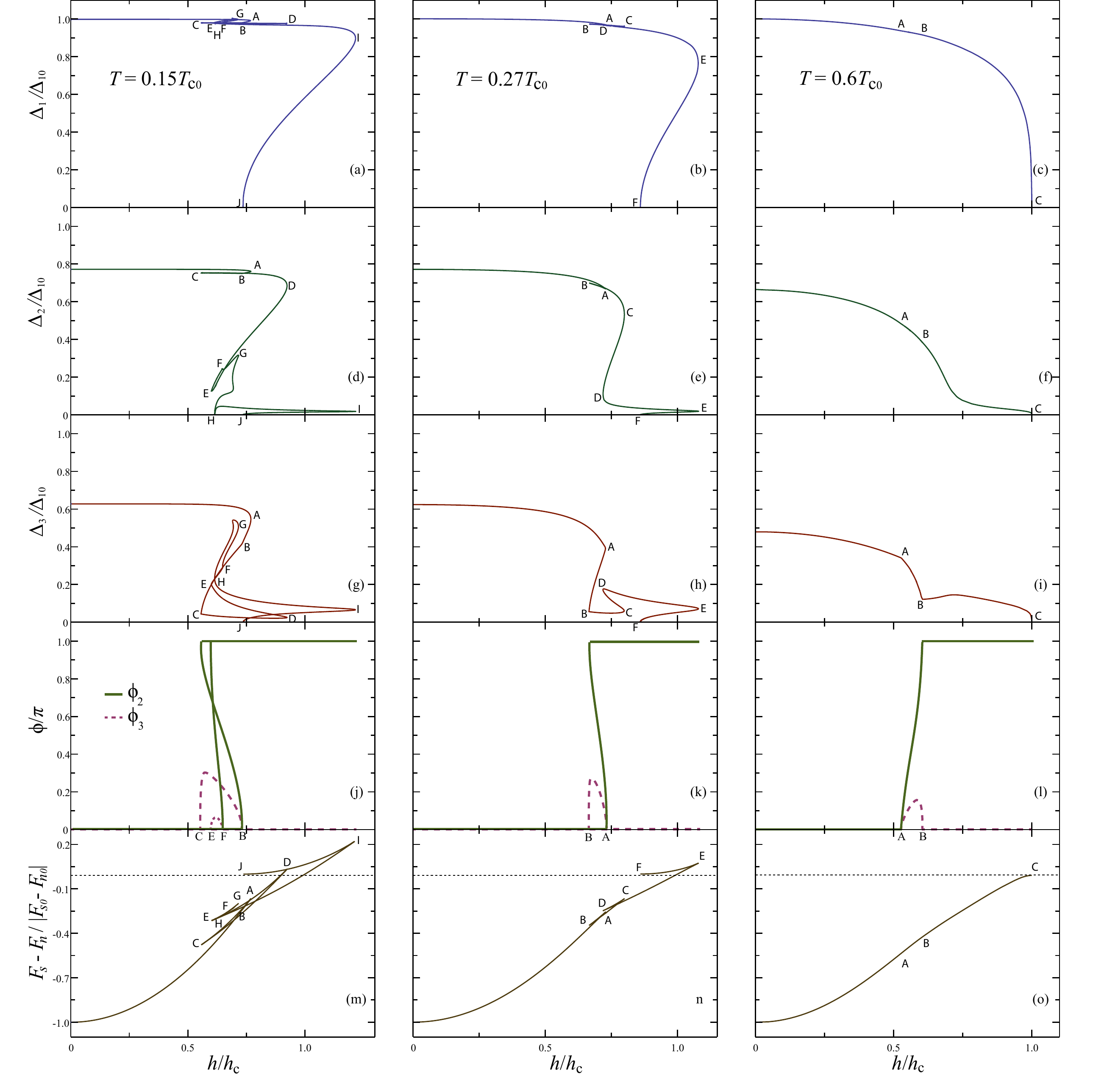}
\end{center}
\caption{Solutions for $\Delta_1$, $\Delta_2$, $\Delta_3$, $\phi_2$, $\phi_3$ and $\Delta F=F_s - F_n$ for three situations, one in each column: $T=0.15 T_{c0}$, $T=0.27 T_{c0}$ and $T=0.6 T_{c0}$. The gap functions are all normalized by $\Delta_{10}$, the value for $\Delta_1$ at zero field for each case. The free energy difference is normalized by its absolute value at zero field and the field $h$ by the thermodynamic critical magnetic field $h_c$ at the corresponding temperature. $\phi_1$ is set to zero as mentioned in Section \ref{sec:af}. The letters are guides to the eye for the reader to follow the continuous path and also mark some points of interest shown in Fig.\ref{h_t_phase_diagram_1}; in the case of the phases, only the letters that mark transitions to or from chiral regions are indicated for $\phi_3$. Dashed lines in the free energy graphics indicate the zero value.}
\label{deltas}
\end{figure*}
The dependence of the gap functions, the superconducting phases and the free energy difference between the superconducting and normal states on applied in-plane magnetic field is shown in Fig.~\ref{deltas}. The temperatures considered for each of the three cases coincide with those of the vertical dashed lines of Fig.~\ref{h_t_phase_diagram_1}. The reader is encouraged to cross data between figures, with particular emphasis on the correspondence of points labeled with letters. The magnetic field evolution of the parameters for a very low temperature, $T=0.15T_{c0}$, where the two regions of chirality of Fig.~\ref{h_t_phase_diagram_1} are present, is shown in the left column in Fig.~\ref{deltas}. In the middle column of Fig.~\ref{deltas}, the temperature, $T=0.27T_{c0}$, is still low enough for the FOTs in the superconducting state to be present, however, it is sufficiently high so as to be outside the second region of chirality shown in the inset of Fig.~\ref{h_t_phase_diagram_1}. The right column of Fig.~\ref{deltas} has $T=0.6T_{c0}$, which is well above the temperature ($T\approx 0.37T_{c0}$) where FOTs in the superconducting state start to occur in Fig.~\ref{h_t_phase_diagram_1}. In the free energy difference plots of Figs.~\ref{deltas}(m),(n) and (o), crossings indicate FOT points and the intersection with the $F_s-F_n=0$ dashed line indicates the superconducting-to-normal state transition $h_c$.

Inspection of the phases of the two higher temperature cases reveals [see Figs.~\ref{deltas}(k) and (l)] the inversion of the magnetic field limits for the chiral phase configurations. If, for $T=0.6T_{c0}$, the system effectively crosses through the chiral region [see section AB in Figs.~\ref{deltas}(i) and (l)], for instance, the same can not be said about $T=0.27 T_{c0}$ where chiral configurations exist only in the metastable section AB [see Figs.~\ref{deltas}(h) and (k)]. In fact, for $T=0.27 T_{c0}$ the system skips over this region due to the FOT a little before point A [in a similar manner to that of the FOT shown in Fig.~\ref{figures7}(h)] and there is a discontinuity in $\phi_2$, that goes from 0 to $\pi$. Therefore, chiral solutions that appear at low temperatures ($T<0.37T_{c0}$) in Fig.~\ref{h_t_phase_diagram_1} simply correspond to metastable solutions. This region of chiral metastable states is shown in a superposition with the non-chiral stable phase configurations in the phase diagram of Fig.~\ref{h_t_phase_diagram_1}. When $T=0.15T_{c0}$ (left column of Fig.~\ref{deltas}), the behavior of $\Delta_2$ and $\Delta_3$ becomes quite complex, with additional reentrances caused mainly by the appearance of a second smaller region of chirality, clearly visible in the phases plots, and occurring in the EF region of the left column of Fig.\ref{deltas}. Again, this second region of chirality does not correspond to stable solutions for the superconducting phases. In fact, they are metastable solutions which occur inside the larger, and also metastable at these low temperatures, chiral region with the diagonal pattern with negative slope.

When we discussed two-band superconductors in section \ref{sec:intro}, we pointed out that there is a large reduction of the gap function of the band with smaller intraband coupling when the FOT within the superconducting phase occurs \cite{Dias2005}. The same kind of behavior can be found twice in our three-band system, on account of there being now two FOTs in the superconducting phase, that is, when the first FOT occurs (red $h_1$ curve in Fig.~\ref{h_t_phase_diagram_1}), $\Delta_3$ experiences a great reduction while both $\Delta_1$ and $\Delta_2$ remain almost constant. When the second FOT occurs (green $h_2$ curve in Fig.~\ref{h_t_phase_diagram_1}), $\Delta_2$ is greatly reduced while again little influence is felt on $\Delta_1$, but $\Delta_3$ actually increases a little. When one reaches the transition given by the blue $h_c$ curve in Fig.~\ref{h_t_phase_diagram_1} by increasing the magnetic field even further, whether it is of first-order for lower temperatures or of second-order for higher temperatures, all three bands undergo a simultaneous transition to the normal phase region of the diagram.

Above the tricritical temperature ($T^* \approx 0.56 T_{c0}$ in Fig.~\ref{h_t_phase_diagram_1}), the dependence of the superconducting phases and gap functions on magnetic field becomes similar to the dependence on temperature, as one can observe in the last column of Fig.~\ref{deltas}.
There is however a steeper  magnetic field dependence  of $\Delta_1$ near $h_c$ which reflects the fact that the system is close to $T^\star$, below which the second-order transition curve becomes of first-order, leading to a reentrant behavior of $\Delta_i(h)$ \cite{Dias2005}.

\subsection{Intermediate interband couplings}\label{sec:strong}

\begin{figure*}[t]
\begin{center}
\includegraphics[width=18.2cm,height=17cm]{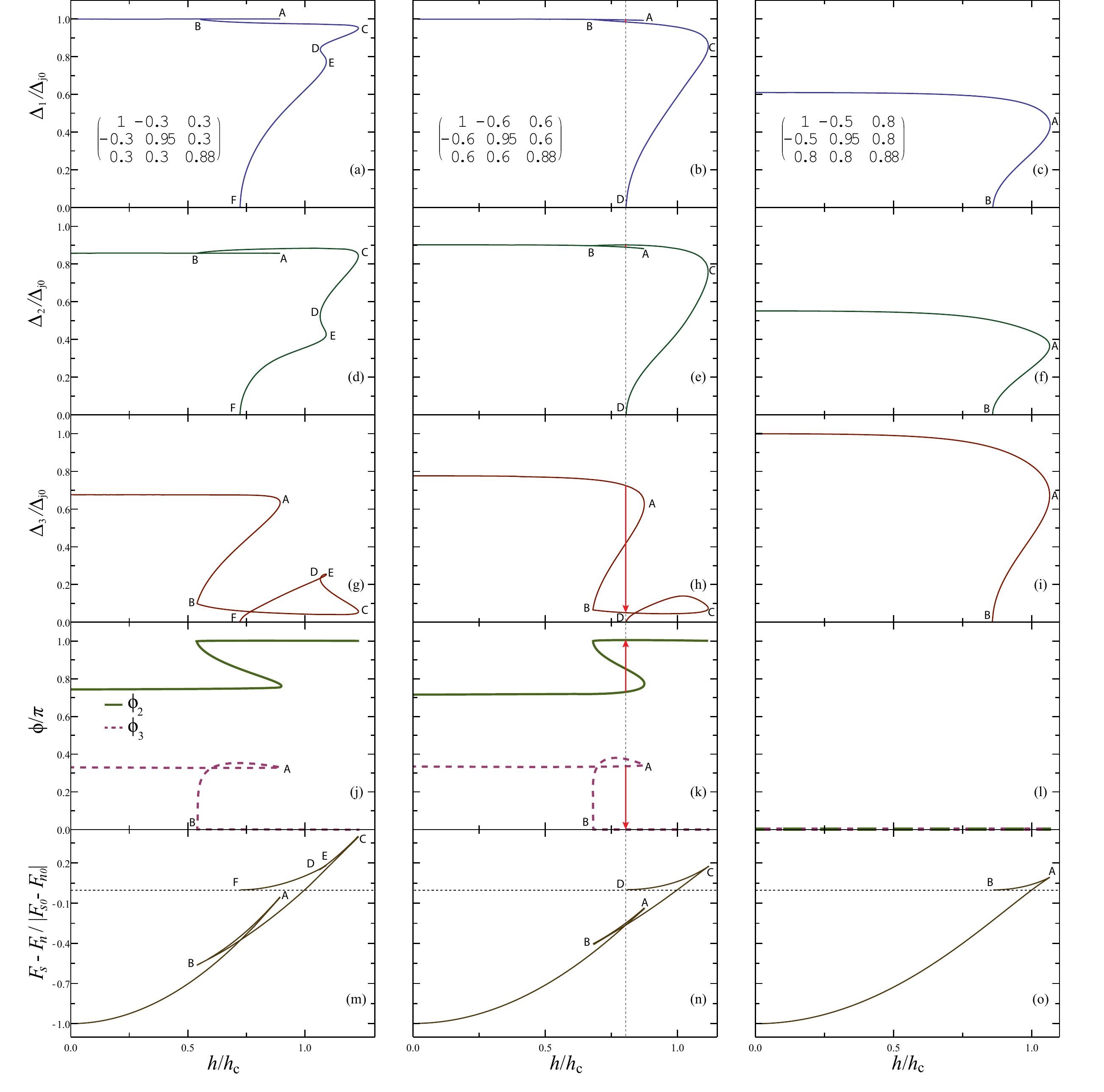}
\end{center}
\caption{Solutions for $\Delta_1$, $\Delta_2$, $\Delta_3$, $\phi_2$, $\phi_3$ and $\Delta F=F_s - F_n$ for three different systems, all at $T=0.2 T_{c0}$, and whose couplings matrices, normalized by $V_{11}$, are given in the top of each column. The gap functions are all normalized by $\Delta_{j0}$, the value for $\Delta_j$ at zero field for each case, with $j=1$ for the two first columns and $j=3$ for the last one. The rest of the notation is the same as in Fig.~\ref{deltas}. The letters are guides to the eye for the reader to follow the continuous path. The arrows in the vertical dashed line of the middle column give the discontinuities in the gap functions (very small for $\Delta_1$ and $\Delta_2$) and phases at the FOT in the superconducting state. Both phases are zero everywhere in (l).} 
\label{figures7}
\end{figure*}

In this subsection we analyze the behavior of three different systems with intermediate interband couplings, all at the same temperature, chosen to be relatively low ($T=0.2 T_{c0}$, where $T_{c0}$ is the critical temperature of the correspondent system, at zero magnetic field) and keeping the same intraband interactions as in the previous section. The cases considered were the following [the potentials, as in (\ref{eq:potential_matrix}), are normalized to $V_{11}$]:
\begin{eqnarray}
1&\to&
\begin{pmatrix}
1 & -0.3 & 0.3
\\
-0.3 & 0.95 & 0.3
\\
0.3 & 0.3 & 0.88
\end{pmatrix}
,\ \  
\\
2&\to&
\begin{pmatrix}
1 & -0.6 & 0.6
\\
-0.6 & 0.95 & 0.6
\\
0.6 & 0.6 & 0.88
\end{pmatrix}
,\ \ 
\\
3&\to&
\begin{pmatrix}
1 & -0.5 & 0.8
\\
-0.5 & 0.95 & 0.8
\\
0.8 & 0.8 & 0.88
\end{pmatrix}.
\end{eqnarray}
The results are shown in Fig.~\ref{figures7}. In case 1, the number of FOTs is reduced to two, instead of the three that are present (at the same temperature) in the weakly coupled bands of Fig.~\ref{h_t_phase_diagram_1}. The transition relative to the second band is absent, but its metastability region is still present in the DE section (see plots in the left column of Fig.~\ref{figures7}), appearing above the superconducting-to-normal state transition at $h_c$, inside its larger metastability region. The system is in a chiral configuration from zero magnetic field and remains that way with increasing magnetic field until the single FOT in the superconducting state is reached, where there is a jump to a non-chiral configuration. By doubling the interband potentials (case 2, middle column of Fig.~\ref{figures7}) not much is altered, but now there is no metastable DE section as in case 1 and the chiral region is extended a little further. For case 2, we indicated explicitly the discontinuities of the gap curves and of the phases at the FOT (red arrows on the dashed vertical line of the middle column of Fig.~\ref{figures7}), where we can see in Fig.~\ref{figures7}(k) that the phases jump from a chiral configuration, $\phi_2,\phi_3\neq 0\vee \pi$, to the non-chiral $(\phi_1,\phi_2,\phi_3)=(0,\pi,0)$ configuration. In the gap curves only the discontinuity at $\Delta_3$ is perceptible [see Fig.~\ref{figures7}(h)], whereas in $\Delta_1$ and $\Delta_2$ [see Figs.~\ref{figures7}(b) and (e)] the jumps are very small. Following the curve of minimum values of the free energy difference, one will get a FOT in the superconducting state ($F_s-F_n<0$ region) whenever there are crossings in this curve.

In case 3 (right column of Fig.\ref{figures7}), the bands are globally more strongly coupled than in case 2 (but $V_{12}$ is slightly lower in magnitude). Given that the superconducting phases start off in a non-chiral configuration, $(\phi_1,\phi_2,\phi_3)=(0,0,0)$ for $h=0$, and that the ratio between the gap functions [and therefore between the $J_{ij}$ effective Josephson couplings in (\ref{eq:minimization1}) that determine the solutions for the phase configurations] is kept almost constant with increasing field, there is no region of chirality anywhere in the entire $[0,h_c]$ domain. Additionally, only one FOT is now present and that is the global transition to the normal phase region, as can be seen in Fig.~\ref{figures7}(o). It also becomes apparent in case 3 that the dominant band is not univocally determined by the largest intraband potential, since the third band becomes the dominant one even though $V_{33}<V_{22}<V_{11}$, on account of being more strongly coupled to bands 1 and 2 than the latter to each other ($V_{13}=V_{23}>\vert V_{12}\vert$). Case 3 is not comparable to the previous ones since one of the interband couplings ($V_{12}=-0.5$) is considerably smaller than the others ($V_{13}=V_{23}=0.8$). This has a deep influence on the interplay between bands, causing them, for instance, to exhibit no FOTs in the superconducting region at $T=0.2T_{c0}$. Asymmetries in the magnitudes of the interband couplings are expected to favor non-chiral arrangements between the superconducting phases, as seen in case 3. One can make an analogy with a system of three classical XY $1/2$-spins coupled antiferromagnetically: if one of the couplings between spins is sufficiently lower (or higher) than the other two, the spins configuration becomes non-chiral as well. 

When we compare the $h$~\textit{vs.}~$T$ phase diagrams of Figs.~\ref{h_t_phase_diagram_1} and \ref{h_t_phase_diagram_2} with cases 1 and 2 of Fig.~\ref{figures7}, the general feature of having the FOTs in the superconducting region occuring  nearer to the critical field $h_c$ as the interband couplings are increased is observed. In Fig.~\ref{h_t_phase_diagram_2} the only FOT in the superconducting region, red $h_1$ curve, occurs very close to $h_c$. However, it is still present, even though the interband couplings are an order of magnitude greater than those of cases 1 and 2, suggesting that the rate at which the metastable region of the FOTs is shifted upwards in the phase diagram is not linearly related to the increasing interband couplings.


\section{Conclusion}\label{sec:conclusion}

In a superconductor with three or more bands and an odd number of repulsive interband interactions, the relative superconducting phases configuration is determined by the magnitude of the effective interband Josephson couplings involved, which can be varied by two controllable external parameters: temperature and magnetic field. 
When in-plane magnetic fields are applied to multiband superconductors with reduced dimensionality, such as iron-based superconductors, one may neglect orbital magnetic effects and consider only the dominant Zeeman splitting term in the Hamiltonian as a first approximation. 

In this paper, we considered quasi-2D three-band superconductors with different interband couplings and characterized the evolution of the chiral superconducting phases with an applied in-plane magnetic field. For weakly coupled bands, we found a narrow strip of chiral superconducting phases configurations in the in-plane magnetic field versus temperature phase diagram that are only stable for high temperatures, when first-order transitions within the superconducting state are absent. The point where the first first-order transition starts to occur gives also the point where the magnetic field limits of the chiral region are inverted, making chiral configurations fall into the metastable region of this transition for low temperatures.
As the interband couplings are equally increased the chiral region of the phase diagram gets quickly broadened, occupying almost the entire superconducting region of the phase diagram for the strongly coupled bands of Fig.~\ref{h_t_phase_diagram_2}. Another consequence of increasing the interband couplings is that the first-order transitions in the superconducting state start to occur closer to the final transition given by the thermodynamic critical magnetic field, vanishing after they intersect. However, the fact that there is still one first-order transition in the superconducting state for the strongly coupled bands of Fig.~\ref{h_t_phase_diagram_2} indicates that it approaches very slowly the final superconducting-to-normal state transition as the interband couplings are increased.

The interplay between the chiral region and the first-order transitions in the phase diagrams of Fig.~\ref{h_t_phase_diagram_1} and \ref{h_t_phase_diagram_2} determines the magnetic field evolution of the superconducting phases configuration at low temperatures. It is also shown that each first-order transition in the superconducting state is characterized by a large reduction in one of the gap functions, while little effect is felt by the other two. 

In systems with intermediate interband couplings we studied the effects of imposing an asymmetry in the magnitude of the interband couplings. From the results found, we argue that, in general, asymmetries of this kind for strong enough interband couplings favor non-chiral configurations for the superconducting phases and tend to suppress first-order transitions in the superconducting state.

Throughout this paper we only considered the Zeeman splitting term as the effect of the in-plane magnetic field. However, even for in-plane magnetic fields the orbital effect is expected to be relevant near the critical temperature, where the Ginzburg-Landau theory is valid. Around the critical temperature a different approach to these frustrated multiband superconductors is required \cite{Yerin2013,Takahashi2014}.

Recalling the equivalence, stated in the introduction, between interband couplings and Josephson junctions \cite{Leggett1966,Agterberg2002,Moll2014}, our studies should also be relevant when addressing circuits of Josephson junctions, particularly when two competing sources of frustration are considered: extrinsic frustration that arises from the application of a magnetic field \cite{Doucot2002,Protopopov2004,Protopopov2006,Pop2008,Pop2010}, and intrinsic frustration that comes from from the insertion of, for example, sign-reversed two-band superconductors or, equivalently, $\pi$-junctions at specific positions \cite{Dias2014}.  

The presence of shoulders in the electronic specific heat below $T_c$ in superconducting materials is one of the ways by which we can infer the existence of multiple bands \cite{Zehetmayer2013}. These shoulders are a consequence of the sharp decrease of the superconducting gaps of the weaker bands around their uncoupled critical temperatures. Discontinuous slope changes in the profiles of the superconducting gaps as functions of the magnetic field,  or of temperature \cite{Dias2011}, when one enters/leaves regions of chiral superconducting phase configurations [see, for example, points A and B in Fig.~\ref{deltas}(i)] should translate, in principle, in the appearance of small peaks or kinks  in the electronic specific heat. These should be present in the proximity of the shoulders generated by
the bands with the smaller gaps, for weakly coupled bands.

The existence of a chiral region in the phase diagram can also be inferred from the behavior of the density of states if non-magnetic impurities are present. When the three bands have equal superconducting phases, non-magnetic impurities are  not pair breaking. In contrast, if one of the three
gaps has an opposite sign to the other two, one expects the same non-magnetic impurities to be pair breaking (as in the case of a  two-band superconductor in a $s\pm$ state \cite{Senga2009,Ng2009}), with the respective appearance of  finite density of states at the Fermi level. Recent experimental results show the existence of in-gap states in iron-based superconductors with non-magnetic impurities \cite{Zhang2014,Mizukami2014}, which seem to be a direct manifestation of a sign reversal in the order parameters between neighboring electron and hole-like bands. By applying a constant in-plane magnetic field to a three-band superconductor and probing the temperature evolution of the system (corresponding, for example, to moving in an horizontal line in Fig.~\ref{h_t_phase_diagram_1}), a continuous modification of the density of states at the Fermi level, due to the appearance of in-gap impurity states, should be observed as one crosses the chiral region of the phase diagram.

One also expects that the critical current in a JJ between a three-band superconductor described in this paper  and a one-band superconductor should reflect the phases changes described above, if the JJ is in the very weak link limit (so that the  phases of the three-band superconductor are determined by the interband couplings in a first approximation). The behavior of the critical Josephson current with temperature or magnetic field should be  similar to the 0-$\pi$ crossover observed in the case  of superconductor-ferromagnet-superconductor Josephson tunnel junction \cite{Weides2006}. That is, the critical Josephson current should continuously decrease as one crosses the chiral region of the phase diagram due to increasing magnetic field or  temperature, reflecting the continuous phase change from a $(0,0,0)$  to a $(0,0,\pi)$ phase configuration.

Several groups have reported evidence supporting the existence of three distinct superconducting gaps in some iron-based compounds, such as $Ba_{1-x}K_xFe_2As_2$ \cite{Storey2013} and $Ba(Fe_{1-x}Co_x)_2As_2$ \cite{Kim2010,Aleshchenko2012,Karakozov2014}. These compounds are good candidates for the observation of the chiral effects we predict here.

\section{Acknowledgments}\label{sec:acknowledments}
R. G. D. acknowledges the financial support from the Portuguese Science and Technology Foundation (FCT) through the program PEst-C/CTM/LA0025/2013.
A. M. M. acknowledges the financial support from the Portuguese Science and Technology Foundation (FCT) through the grant SFRH/BI/52520/2014.
M.A.N.A. acknowledges support from
the Portuguese Science and Technology Foundation (FCT) through Project EXPL/FIS-NAN/1728/2013.

\appendix
\section{Free energy difference between superconducting and normal states}
\label{sec:appendix}
In order to derive expression (\ref{free_energy}) for the free energy difference between the superconducting and normal states of a multiband quasi-2D-superconductor with an in-plane applied magnetic field, we start again by considering the Hamiltonian for $n$-bands with an extra Zeeman term given by (\ref{eq:hamiltonian}).
Assuming a BCS ground-state extended to multiple bands of the form
\begin{equation}
\vert GS \rangle = \prod_{\textbf{k}j}(u_{\textbf{k}j} + e^{i\phi_j}v_{\textbf{k}j} c_{\textbf{k}\uparrow j}^\dag c_{-\textbf{k}\downarrow j}^\dag)\vert 0 \rangle,
\end{equation}
where we assigned an overall superconducting phase on band $j$ to the coefficient $v_{kj}$, keeping $u_{kj}$ real, with no loss of generality. Using the standard Bogoliubov-Valatin transformations
\begin{eqnarray}
\begin{pmatrix}
\gamma^\dagger_{k\uparrow j}
\\
\gamma_{-k\downarrow j}
\end{pmatrix}
&=&
\begin{pmatrix}
u_{kj} & -v_{kj}^\star
\\
v_{kj} & u_{kj}
\end{pmatrix}
\begin{pmatrix}
c^\dagger_{k\uparrow j} 
\\
c_{-k\downarrow j}
\end{pmatrix} ,
\\
\Big\{\gamma^\dagger_{k\uparrow i}, \gamma_{k^\prime\uparrow j}\Big\}&=&\delta_{kk^\prime}\delta_{ij} ,
\end{eqnarray}
and the expression for the thermal average of an operator
\begin{equation}
    \langle \hat{O} \rangle=
    \frac{Tr(\hat{O}e^{-\beta\mathcal{H}})}{Tr(e^{-\beta\mathcal{H}})} ,
\end{equation}
the mean-field approximation leads to  a new expression for the Hamiltonian as the sum of a term $\mathcal{H}_0$ for the independent quasi-particle excitations and a constant term $c$ representing the thermal average of the Hamiltonian,
\begin{equation}
\mathcal{H}=\mathcal{H}_0 + c ,
\end{equation}
with
\begin{eqnarray}
\mathcal{H}_0 &=& \sum_{\textbf{k}\sigma i}E_{\textbf{k}i}^\sigma \gamma_{\textbf{k}\sigma i}^\dagger \gamma_{\textbf{k}\sigma i} ,
\\
E_{ki}^\sigma &=& \sqrt{\xi_{ki}^2 + \vert\Delta_{ki}\vert^2} + \sigma h,
\end{eqnarray}
where $E_{ki}^\sigma$ gives the excitation spectrum, and 
\begin{widetext}
\begin{eqnarray}
    c
    &=&
    \sum_{\textbf{k}\sigma i}\bigg\{v_{\textbf{k}i}^2\xi_{\textbf{k}i} 
    + 
    \Big[\xi_{\textbf{k}i}(u_{\textbf{k}i}^2-v_{\textbf{k}i}^2) - E_{\textbf{k}i} \Big]f(E_{\textbf{k}i}^\sigma)\bigg\}
    \\ 
    &-& 
    \sum_{\textbf{kk}^\prime i}V_{\textbf{kk}^\prime}^{ii} 
    u_{\textbf{k}i}v_{\textbf{k}i}u_{\textbf{k}^\prime i}v_{\textbf{k}^\prime i}
    \big(1-f(E_{\textbf{k}i}^\uparrow)-f(E_{\textbf{k}i}^\downarrow)\big) 
    \big(1-f(E_{\textbf{k}^\prime i}^\uparrow)-f(E_{\textbf{k}^\prime i}^\downarrow)\big)           
    \nonumber
    \\
    &-&
    2 \sum_{\textbf{kk}^\prime , j>i}\cos(\phi_j - \phi_i) 
    V_{\textbf{kk}^\prime}^{ij} u_{\textbf{k}i}v_{\textbf{k}i}u_{\textbf{k}^\prime j}v_{\textbf{k}^\prime j}
    \big(1-f(E_{\textbf{k}i}^\uparrow)-f(E_{\textbf{k}i}^\downarrow)\big) 
    \big(1-f(E_{\textbf{k}^\prime j}^\uparrow)-f(E_{\textbf{k}^\prime j}^\downarrow)\big),
    \nonumber
\end{eqnarray}
\end{widetext}
where $f(E_{ki}^\sigma)=\langle \gamma_{ki\sigma}^\dagger \gamma_{ki\sigma} \rangle$ is the ideal fermi gas occupation number and $E_{ki}=E_{ki}^\sigma+\sigma h$. Note that only phase differences between bands in the interband term have any physical meaning and not the absolute values of the phases themselves, which means that in a system with $i$ bands we only need to determine $i-1$ phase differences. The free energy will be given by
\begin{eqnarray}
F &=& -k_BT\ln\big[Z_{red}\big]
\\
&=& -\frac{1}{\beta}\ln\big[Tr(e^{-\beta(\mathcal{H}_0+c)})\big] \nonumber
\\
&=& -\frac{1}{\beta}\sum_{\textbf{k}\sigma i}\ln\big[1+e^{-\beta E_{\textbf{k}\sigma i}}\big] + c \nonumber
\\
&=& k_BT\sum_{\textbf{k}\sigma i}\ln\big[1-f(E_{\textbf{k}\sigma i})\big] + c . \nonumber
\end{eqnarray}
We define now the quantities
\begin{eqnarray}
\Delta_i &=& \sum_j \cos(\phi_j - \phi_i)V_{ij}\delta_j,
\label{eq:Deltas1}
\\
\delta_j &=& \sum_k u_{kj}v_{kj}\Big(1-f(E_{kj}^\uparrow)-f(E_{kj}^\downarrow)\Big) ,
\label{eq:deltas1}
\end{eqnarray}
where the BCS approximations $\Delta_{ki},\delta_{ki},V_{kk^\prime}^{ij} \to \Delta_i,\delta_i,V_{ij}$ are implied (s-wave symmetry is assumed).  Note that (\ref{eq:Deltas}) and (\ref{eq:deltas}) are just the integral versions of (\ref{eq:Deltas1}) and (\ref{eq:deltas1}), respectively, over an energy interval given by the Debye frequency.  Using the identity
\begin{equation}
u_{ki}^2 + v_{ki}^2=1
\end{equation}
and the conditions resulting from the minimization of the free energy with respect to $v_{ki}$, 
\begin{eqnarray}
\Delta_i^2&=&E_{ki}^2-\xi_{ki}^2 ,
\\
v_{ki}^2&=&\frac{1}{2}\big(1-\frac{\xi_{ki}}{E_{ki}}\big) ,
\\
u_{ki}v_{ki}&=&\frac{\Delta_i}{2E_{ki}} ,
\\
u_{ki}^2-v_{ki}^2&=&\frac{\xi_{ki}}{E_{ki}} ,
\end{eqnarray}
one has
\begin{eqnarray}
\delta_i - \sum_\textbf{k}\frac{\Delta_i}{2E_{\textbf{k}i}} &=& -\sum_\textbf{k}\frac{\Delta_i}{2E_{\textbf{k}i}}\Big[f(E_{\textbf{k}i}^\uparrow)+f(E_{\textbf{k}i}^\downarrow)\Big] ,
\end{eqnarray}
and
we end up with
\begin{equation}
F_s=k_BT\sum_{\textbf{k}\sigma i}\ln\big[1-f(E_{\textbf{k}i}^\sigma)\big] + \sum_{\textbf{k}i}(\xi_{\textbf{k}i}-E_{\textbf{k}i}) + \sum_i \delta_i\Delta_i .
\end{equation}

The normal state free energy corresponds to the particular case $F_n=F_s(\Delta_i=0)$ or, conversely, $E_{ki}^\sigma \to \vert \xi_{ki}^\sigma \vert=\vert \xi_{ki}\vert - \sigma h$ 

\begin{equation}
F_n=k_BT\sum_{\textbf{k}\sigma i}\ln\big[1-f(\vert \xi_{\textbf{k}i}^\sigma \vert)\big] + 2\sum_{\vert \textbf{k} \vert < k_F}\xi_{\textbf{k}i},
\end{equation}
where $\mu=\varepsilon_F$ is assumed. Finally we have everything we need to get the expression for the free energy difference between states,
\begin{eqnarray}
\Delta F &=& F_s - F_n 
\\
&=&k_BT\sum_{\textbf{k}\sigma i}\ln\bigg[\frac{1-f(E_{\textbf{k}i}^\sigma)}{1-f(\vert \xi_{\textbf{k}i}^\sigma \vert)}\bigg] + \sum_{\textbf{k}i}(\xi_{\textbf{k}i}-E_{\textbf{k}i}) \nonumber
\\
&-& 2\sum_{\vert \textbf{k} \vert < k_F}\xi_{\textbf{k}i} + \sum_i\delta_i\Delta_i
\nonumber
\\
&=&k_BT\sum_{\textbf{k}\sigma i}\ln\bigg[\frac{1-f(E_{\textbf{k}i}^\sigma)}{1-f(\vert \xi_{\textbf{k}i}^\sigma \vert)}\bigg] + 2\sum_{\vert \textbf{k} \vert > k_F,i}(\xi_{\textbf{k}i}-E_{\textbf{k}i}) \nonumber \\
&+&  \sum_i\delta_i\Delta_i . \nonumber
\end{eqnarray}
To find the free energy difference profiles in Figs. \ref{deltas} and \ref{figures7} we used this equation in the continuum limit where the first two sums were turned to integrals limited by the Debye frequency $\omega_D$, with the approximation $\frac{\hbar\omega_D}{\Delta_i}\gg 1$.

\bibliography{frustrated}

\end{document}